\documentclass[journal=jacsat,manuscript=article]{achemso}

\usepackage{chemformula} 
\usepackage[T1]{fontenc} 

\newif\ifComment
\Commentfalse
\graphicspath{{./Figs/}}

 \SectionNumbersOn
\author{Emadeldeen Hassan}
\email{emadeldeen.hassan@umu.se}
\affiliation[UMU]
{{Department of Applied Physics and Electronics, Ume\aa \mbox{} University, SE-901~87~Ume{\aa}, Sweden}}
\alsoaffiliation[MNF]
{Department of Electronics and Electrical Communications, Menoufia University, Menouf 32952, Egypt}

\author{Andrey B. Evlyukhin}
\affiliation[LUHAE]{Institute of Quantum Optics, Leibniz University Hannover, 30167 Hannover, Germany} 
\alsoaffiliation[PhoenixD]{Cluster of Excellence PhoenixD, Leibniz University Hannover, 30167 Hannover, Germany} 

\email{evlyukhin@iqo.uni-hannover.de}

\author{ Antonio Cal\`a Lesina}
\affiliation[LUHAC]{Hannover Centre for Optical Technologies, Leibniz University Hannover, 30167 Hannover, Germany}
\alsoaffiliation[PhoenixD] {Cluster of Excellence PhoenixD, Leibniz University Hannover, 30167 Hannover, Germany}
\alsoaffiliation[ITAAC]{Institute for Transport and Automation Technology, Faculty of Mechanical Engineering, Leibniz University Hannover, 30167 Hannover, Germany}

\email{antonio.calalesina@hot.uni-hannover.de}

\title[An \textsf{achemso} demo]
  {Anapole plasmonic meta-atom enabled by inverse design for metamaterials transparency}

\abbreviations{IR,NMR,UV}

\begin{document}


	
\begin{abstract}
Anapole states are broadly investigated in nanophotonics for their ability to provide field enhancement and transparency. 
While low extinction has been achieved in dielectric nanoparticles due to the absence of intrinsic losses, in the case of plasmonic nanostructures this is still lacking. 
Here, we report an easy-to-fabricate planar plasmonic nanostructure found via topology optimization, which exhibits an anapole state with close-to-ideal characteristics in the visible regime including weak absorption, high near-field enhancment, and strong suppression of scattering.
The nanonantenna can act as an individual meta-atom because, due to low inter-coupling, it preserves its optical response even when used in highly packed metasurfaces and metamaterials. 
The low losses are due to the optimized topology which provides concentration of the field outside the structure with minimum penetration inside the metal. Compared to anapole states in dielectric structures, the accessibility of the volume of enhanced field is suitable for sensing applications.
Anapole states are typically interpreted as the result of the interference between electric and toroidal dipole moments. 
Here and based on a novel approach, in the context of secondary multipole analysis, we introduce the concept of anapole state without using the contribution of toroidal dipole moments.
The article provides new insight into anapoles in plasmonic nanostructures and ways to achieve them, while remarking the power of topology optimization to unlock designs with novel functionalities.
\end{abstract}


\section{Introduction}
In recent years, nanophotonic structures have attracted much attention for their property to scatter the incoming light in ways that can be engineered by design. 
Exotic nanostructured designs have been proposed that can support charge-current distributions which are able to sustain any arbitrary radiated or non-radiated field.
Nonradiating sources, that exhibit the so-called anapole states, represent a special class of electromagnetic objects, for their ability to provide the seemingly contradictory features of field enhancement in the near-field and zero scattering (i.e., unitary transmittance when used in array configuration) in the far-field \cite{savinov19optical,zanganeh2021anapole}. The first anapole demonstration in the microwave regime was related to a metallic toroidal metamaterial \,\citep{kaelberer10toroidal,Fedotov2013}, while the first case in optics was reported for a high-index dielectric nanodisk\,\cite{miroshnichenko2015nonradiating}. For more details about nonradiating anapole states in nanophotonics, we suggest the review articles\,\cite{Bozhevolnyi2019,Baryshnikova2019,saadabad2022multifaceted}, where historical background on the development of the field is also provided. 

The exact multipole decomposition and anapole identification requires the expansion of the far-field in spherical harmonics\,\cite{Baryshnikova2019}. 
However, for nanostructures much smaller than the wavelength, their Taylor expansion in Cartesian coordinates (known as long-wavelength approximation), which involves the current densities in the nanostructure, provides a better insight about the formation of the anapole state\,\cite{corbaton15Exact,gurvitz2019high}.
Although the most known way to achieve an anapole state is via the destructive interference between toroidal dipole and electric dipole, other multipole terms can also contribute to the anapole creation.
Significative examples are the hybrid anapole\,\citep{CanosValero2021}, and the anapole due to transverse Kerker effect\,\citep{terekhov2019multipole,Shamkhi2019}. 
An extension of the work in\,\citep{CanosValero2021}, was proposed in\,\citep{Kuznetsov2021} where a dielectric nanostructure exhibits an anapole state which makes the nanostructure resemble a perfect meta-atom, with performance that does not depend on neighbors (negligible coupling), thus paving the way to highly packed metasurfaces, also with random distribution of the nanostructures and negligible effect of the substrate. 
Furthermore, active control of anapole states in dielectric metasurfaces via phase-changing materials was discussed in the literature\,\citep{Tian2019}.

The examples reported above are based on dielectrics because anapole states require low losses. 
However, dielectric nanostructures only provide field enhancement inside the structure volume, which is not ideal for sensing and some nonlinear optics applications, for which metals are preferred.
Despite the effort to reduce losses in plasmonic materials\,\citep{Naik2013}, the intrinsic absorption of metals hinders the possibility to achieve a close-to-ideal performance in anapoles, even in numerical studies. 
Despite these limitations, several anapole states in metallic nanostructures have been theoretically investigated, such as in a plasmonic metamaterial composed of judiciously arranged U-shaped split ring resonators\,\citep{Huang2012}, a gap-surface plasmon resonators\,\citep{Yezekyan2022}, and a design that overcomes losses by exploiting a gain material\,\citep{Pan2021}. 
Anapole states were demonstrated in the terahertz regime via dumbbell apertures in a stainless steel sheet\,\citep{Li2021}, and a similar geometry was also adopted in the infrared regime\,\citep{wu2018optical}. 
In the latter case, an array of dumbbell shape apertures on a gold sheet was combined with an array of vertical split ring resonators, which are responsible for the creation of toroidal dipole at each aperture location. However, such system is complex to fabricate as it requires vertical split ring resonators and alignment between two metasurfaces.

In this paper, we address the question whether it is possible to achieve nearly ideal anapole states in plasmonic nanostructures in the visible regime that are easy to fabricate, i.e., planar.
A positive answer is suggested by inverse design via topology optimization\,\citep{Bendsoe2004, WADBRO06Topoloy,Hassan14Topology,Aage17Giga,molesky18inverse,christiansen19nonlinear,Hassan20Multilayer, Hammond22High}. 
The plasmonic nanostructure presented in this paper was found by performing the optimization to maximize the field enhancement in a gap volume in the middle of the structure.
The optimization was performed using an algorithm recently developed in our group\,\citep{Hassan2022}, which allows us to account for Drude dispersion in one run of the code and thus to get a broadband overview of the device performance. 
This allowed us to identify a spectral region with extremely low scattering cross-section, i.e., an anapole state, where the nanostructure also exhibits high-field enhancement in the gap.
It emerges that such anapole has close-to-ideal characteristics, including very low losses due to the optimized topology that localizes the field in the open volume region around the gap with minimum penetration into the metal. Furthermore, our anapole nanostructure acts as an independent plasmonic meta-atom, similar to what is discussed for dielectric nanostructures in\,\citep{Kuznetsov2021}. The nanostructure is analyzed stand-alone, in metasurface and metamaterial configuration, and it preserves its features in all cases, including the case when placed on a substrate, thus enabling the realization of metasurfaces with a high density of plasmonic hot-spots, as well as plasmonic metamaterials with high transparency. This was not observed in \,\citep{wu2018optical} because the unit cells were all connected, which forced the authors to seek the introduction of a toroidal dipole, and prevented them from achieving a planar structure. 
In our case, the planar structure makes the toroidal dipole contribution not essential to the formation of the anapole state. Thus, we refer to the minimum in the electric dipole as the origin of the anapole. By means of a secondary multipole analysis, we explain the anapole effect by only referring to the electric dipole moment. We identify two electric dipole moments with opposite phases, associated with different subvolumes of the nanostructure, which explain the formation of the anapole state due to their destructive interference.
The features of our design are relevant for nonlinear optics enhancement and nonlinear light structuring. In fact, the close vicinity of the nanostructures due to low coupling can be exploited for nonlinear light structuring without diffraction\,\cite{Lesina2017}, and lasing.
In Section\,\ref{Sec:TopOptNan}, we present the optical numerical characterization of the nanostructure that elucidates the anapole state.
In Section\,\ref{Sec:MulAna}, we employ multipole analysis to clarify the mechanism of scattering cancellation.
In Section\,\ref{Sec:MetBuiBlo}, we demonstrate the possibility of achieving transparent metasurfaces, while providing field enhancement. 
In Section\,\ref{Sec:AppMet}, we extend the demonstration to a metamaterial setup.
Conclusion and future perspectives are then presented in Section\,\ref{Sec:Con}.

\section{Topology optimized structure}
\label{Sec:TopOptNan}

\begin{figure}
\centering
\includegraphics[trim = 9mm 17mm 12mm 12mm, clip,width=0.9\textwidth,draft=false,angle=0]{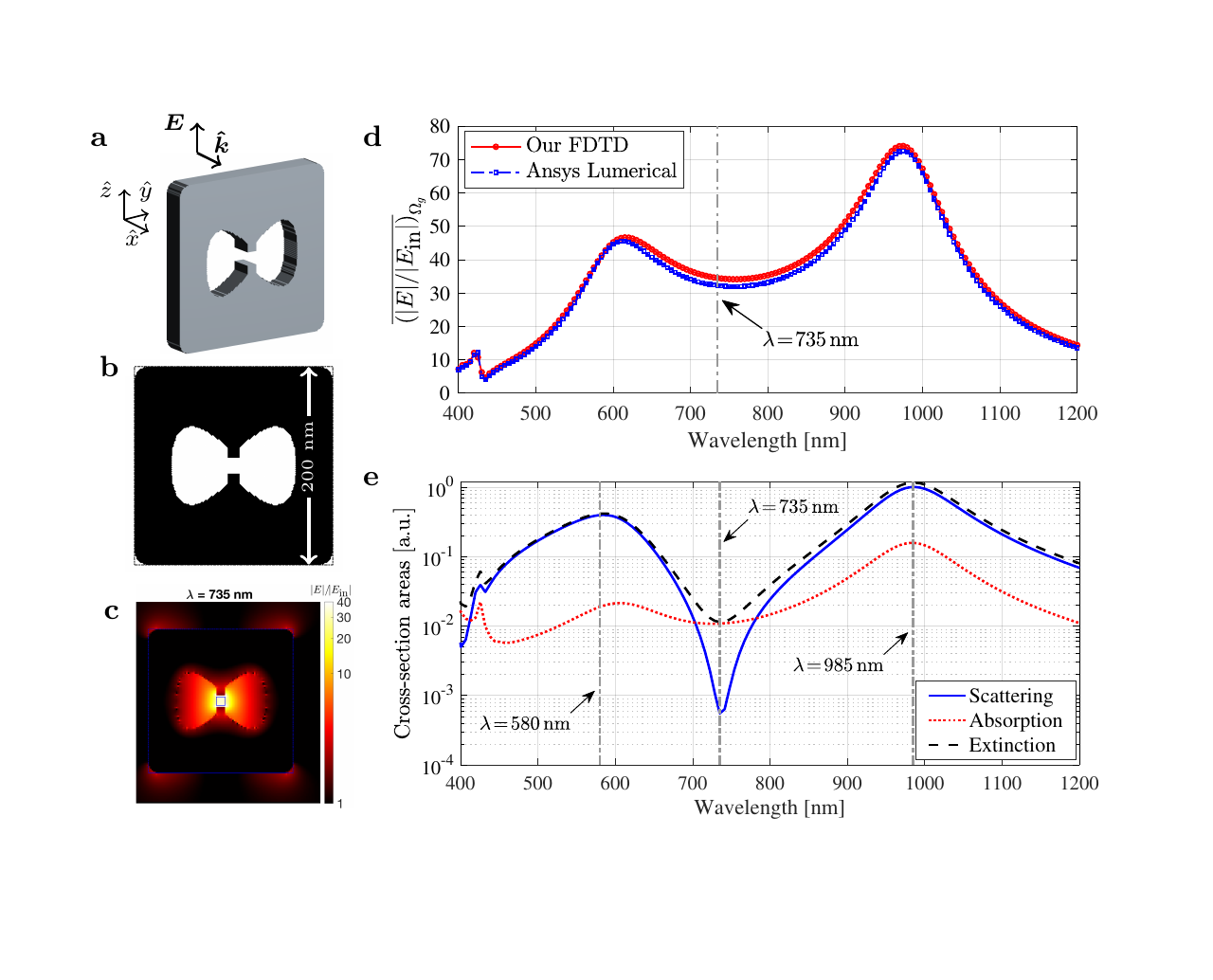}
   \vspace{-0pt}
\caption{Optimized silver nanostructure. (a) 3D view and (b) cross-sectional view of the optimized design. (c) field distribution at the anapole wavelength $\lambda \!=\!735$\,nm, marked by the vertical dash-dotted line in (d) and (e). (d) average field enhancement inside the gap region, located at the center of the nanostrucure, cross-verified with Ansys Lumerical commercial software package\,\cite{Ansys}. (e) scattering, absorption, and extinction cross-sections of the nanostructure revealing an anapole state (computed using Ansys Lumerical).}
  \label{Fig1}
   \vspace{-8pt}
\end{figure}

In a recent work \cite{Hassan2022}, we introduced a new topology optimization method to optimize wideband dispersive plasmonic nanostructures.  
The method distributes a given material in a specified design domain in order to extremize a defined objective function.   
The large degrees of freedom offered by the method enable efficient non-intuitive structures with outstanding performance. 
We applied the method to optimize the topology of a silver nanostructure to maximize the electric field of an incident plane wave into a gap region positioned at the center of the nanostructure.
Figure\,\ref{Fig1}a and Fig.\,\ref{Fig1}b show, respectively, a 3D perspective view and a 2D transverse view of one of the optimized nanostructures, which has the dimensions of $200\times200\times30$\, nm${}^3$. 
The nanostructure has a gap region $\Omega_g$, located at the center of the nanostructure, with dimensions $12\times 12 \times 30$\,nm${}^3$.
The optimized nanostructure exhibits more than $30$-fold average field enhancement over the spectral range $560$--$1180$\,nm, as shown in Fig.\,\ref{Fig1}d, where we cross-verify computations by our FDTD\,code with the commercial software package Ansys Lumerical FDTD\,\cite{Ansys}.
It is worth mentioning that our nanostructure has structural similarities with previously reported ones in the literature, where similar structures have been studied, however, in the context of periodic metasurfaces, including  a metal film perforated by corresponding apertures \cite{yue2014enhanced,wu2018optical}.

The scattering cross-section shown in Fig.\,\ref{Fig1}e reveals two peaks, located around $580$\,nm and $985$\,nm, and one dip located around $735$\,nm, where the scattering cross section decreases significantly by almost four orders of magnitude compared to the peak values.
In the same figure, we show the absorption and extinction cross-sections, which essentially follow the same behaviour as the scattering curve.
That is, although the nanostructure exhibits the near-field enhancement capability at the dip around $735$\,nm, as demonstrated in Fig.\,\ref{Fig1}c, it exhibits a small extinction cross-section caused mainly by its relatively small absorption around $735$\,nm. 
The combined features of the nanostructure at $735$\,nm suggests the presence of an anapole state.   
To understand the origin of such anapole state, we analyze the nanostructure using multipole analysis in the next section.

\section{Multipole analysis}
\label{Sec:MulAna}
In this section, we briefly present the multipole analysis approach. 
The approach is then employed to analyze the nanostructure and find the contributing multipoles to its far-field.
We investigate the nanostructure using the conventional multipole analysis which, for our nanostructure, was not sufficient to explain the origin of the anapole state.
Therefore, we employ a secondary multipole analysis that can explain the exotic features of the nanostructure around $735$\,nm and associate these features to an anapole state.
Unlike what is traditionally thought about achieving the anapole state by the superposition of electric and toroidal dipoles, our secondary multipole analysis reveals a new concept to realize the anapole state mainly based on contributions by quasi-static electric dipole moments. In our approach, we consider that all involved multipole moments are located at the center of symmetry of the nanostructure.

\subsection{Theoretical background}

The multipole expansion of the scattering cross section (SCS), $\sigma_{\rm sca}$, with explicit inclusion of the electric dipole (ED) vectorial moment $\bf p$ and magnetic quadrupole (MQ) tensorial moment $\hat M$ can be written as\,\cite{Evlyukhin_PhysRevB_2016}
\begin{equation}\label{SCS}
\sigma_{\rm sca}=\frac{k_0^4}{6\pi\varepsilon_0^2|{\bf
E}|^2}|{\bf p}|^2
+\frac{k_0^6
\varepsilon_d^2\mu_0}{80\pi\varepsilon_0 |{\bf
E}|^2}\sum_{\alpha\beta} |{ M_{\alpha\beta}}|^2+\cdots,
\end{equation}
where $k_0$ is the wavenumber in vacuum, $\varepsilon_0$ is the vacuum permittivity, $\varepsilon_d$ is the relative permittivity of the surrounding medium, $\mu_0$ is the vacuum permeability, $\bf E$ is the incident electric field at the point of the  multipole  moment localization.  
The integral expressions that determine the exact multipole moments are discussed in the literature\,\cite{alaee2018electromagnetic,Evlyukhin_PhysRevB_2019}, and for the ED and the MQ moments can be presented as
\begin{flalign}\label{p0}
{\bf p}&=\frac{i}{\omega}\int_{V_s}j_0(k_dr'){\bf j}({\bf r}')d{\bf r}'+ \nonumber\\ 
&\quad +\frac{ik_d^2}{{2}\omega}\int_{V_s}\frac{j_2(k_dr')}{(k_dr')^2}\{3[{\bf r}'\cdot{\bf j}({\bf r}')]{\bf r}'-{r}'^2{\bf j}({\bf r}')\}d{\bf r}',
\end{flalign}
\begin{equation}\label{M2}
{\hat M}={5}\int_{V_s}\frac{j_2(k_dr')}{(k_dr')^2}\{[{\bf r}'\times{\bf j}({\bf r}')]{\bf r}'+{\bf r}'[{\bf r}'\times{\bf j}({\bf r}')]\}d{\bf r}',
\end{equation}
where $k_d=k_0\sqrt{\varepsilon_d}$; the vector $\mathbf{j}({\bf r}')$ is the electric current density induced in the scatterer by an incident light wave with the angular frequency $\omega$; $\mathbf{r'}$ is the radius vector of a volume element inside the scatterer with the origin of the coordinates located at the center of the nanostructure; $V_s$ is the scatterer's volume; and $j_n(k_d r')$ is the $n$th order spherical Bessel function. 

Note that in Eq.\,\eqref{SCS}, we explicitly present the main multipole contributions in the SCS of the considered structure.  
Including only these terms provides a good approximation of the full SCS. 
Thus, the SCS is expressed as
\begin{equation}
    \sigma_{\rm sca}\simeq\sigma_{\rm sca}^p+\sigma_{\rm sca}^M,
\end{equation}
where the first and the second terms are the ED and the MQ contributions, respectively.

\subsection{Toroidal dipole and anapole state}
By expanding the Bessel functions in Eq.\,\eqref{p0} in a Taylor series (near zero) and using only the first two terms, the electric dipole moment vector can be approximated by \cite{Evlyukhin_PhysRevB_2016,alaee2018electromagnetic}
\begin{flalign}\label{LWA_p}
    {\bf p}& \approx\frac{i}{\omega}\int_{V_s}{\bf j}({\bf r}')d{\bf r}'+\frac{ik_d^2}{10\:\omega}\int_{V_s}\{[{\bf r}'\cdot{\bf j}({\bf r}')]{\bf r}'-2{r}'^2{\bf j}({\bf r}')\}d{\bf r}' \nonumber \\ 
    & \equiv{\bf p}^0+\frac{ik_d^2}{\omega}{\bf T},
\end{flalign}
where the first integral term, denoted as ${\bf p}^0$, is the long wavelength approximated electric dipole (ED$_0$) moment; the second integral term  includes the toroidal dipole (TD) moment ${\bf T}$. 
We emphasize that expression\,\eqref{LWA_p} corresponds to the Long Wavelength Approximation (LWA) where the argument of the Bessel functions in the multipole definitions is sufficiently small for all points ${\bf r}'$ inside the scatterer \cite{alaee2018electromagnetic}. 
In the LWA, the electric dipole contribution to $\sigma_{\rm scat}$ is 
\begin{equation} \label{p1}
    \sigma_{\rm sca }^p\approx\sigma_{\rm sca }^{{\rm ED}_0+{\rm TD}}=\frac{k_0^4}{6\pi\varepsilon_0^2|{\bf
E}|^2}|{\bf p}^0+\frac{ik_d^2}{\omega}{\bf T}|^2,
\end{equation}
which can be expressed as  
\begin{equation}\label{p2}
    \sigma_{\rm sca}^{{\rm ED}_0+{\rm TD}} \!=\!\frac{k_0^4}{6\pi\varepsilon_0^2| {\bf
E}|^2}\!\left\{\! |{\bf p}^0|^2+\frac{k_d^4}{\omega^2}|{\bf T}|^2 \!-2\frac{k_d^2}{\omega}\Im({\bf T}\cdot{\bf p}^{0*})\! \right\}.
\end{equation}
In the traditional concept of a dynamic anapole \cite{miroshnichenko2015nonradiating,saadabad2022multifaceted}, the suppression of electric dipole scattering (radiation) is associated with  the destructive interference of fields created by the quasi-static electric dipole moment ${\bf p}^0$ and the toroidal dipole moment $\bf T$. 
From Eq.\,\eqref{p1} and Eq.\,\eqref{p2} and in order for $\sigma_{\rm sca}^p\to0$, the following condition should be satisfied:
\begin{equation}
    {\bf p}^0=-\frac{ik^2_d}{\omega}{\bf T} \quad{\rm or}\quad |{\bf p}^0|^2+\frac{k_d^4}{\omega^2}|{\bf T}|^2=2\frac{k_d^2}{\omega}\Im({\bf T}\cdot{\bf p}^{0*}).
\end{equation}
Note that here, in order to realize the anapole state, the excitation of a TD moment of a certain magnitude and phase is required.

\begin{figure}
\centering
\includegraphics[trim = 0mm 0mm 0mm 0mm, clip,width=0.99\textwidth,draft=false,angle=0]{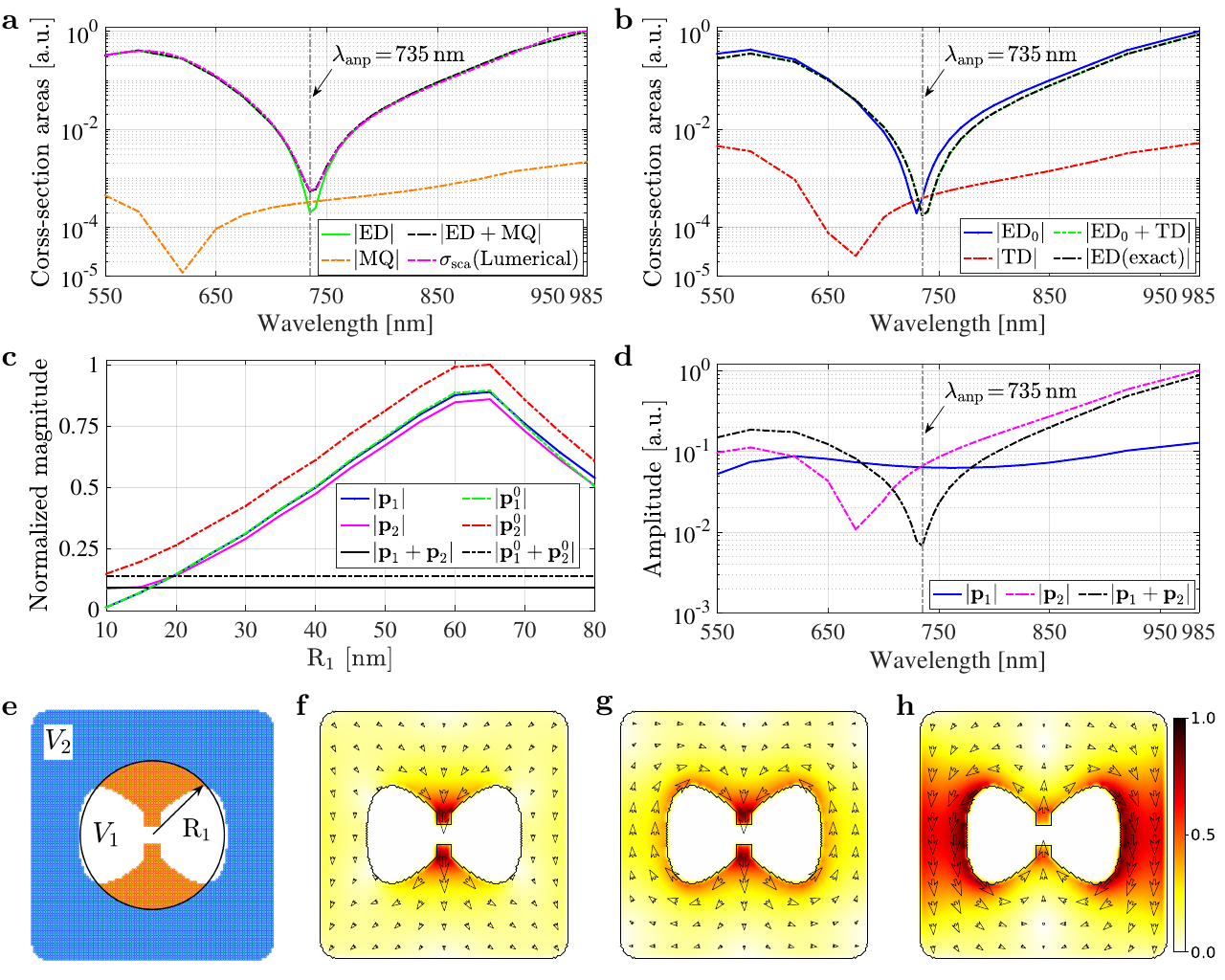}
   \vspace{-0pt}
\caption{Nanostructure multipole analysis. (a) total scattering cross-section and corresponding exact electric dipole (ED) and magnetic quadrupole (MD) contributions. (b) contributions of the quasi-static electric dipole (ED$_0$), toroidal dipole (TD), the sum (ED$_0$+TD), and the exact electric dipole (ED) to the total scattering cross-section. 
(c) absolute values of the exact ED ($\bf{p}=\bf{p}_1+ \bf{p}_2$) and the LWA ED ($\bf{p}^0=\bf{p}^0_1+ \bf{p}^0_2$) of the whole structure, and the individual contributions from sub-volumes $V_1$ and $V_2$, calculated at $\lambda=735$\,nm for different radii R$_1$, see (e). (d) spectral dependency of the absolute values of the ED ${\bf p}_1$ of $V_1$, and the ED ${\bf p}_2$ of $V_2$, calculated for R$_1$=60\,nm, with the total ED $\bf{p} = \bf{p}_1+\bf{p}_2$. (e) splitting the nanostructure using the radius R$_1$ into sub-volumes $V_1$  and $V_2$. 
 (f), (g) and (h) normalized electric field distribution at the center of the nanostructure at the wavelength 580\,nm, 735\,nm, and 985\,nm, respectively.}
  \label{Fig2}
   \vspace{-8pt}
\end{figure}

We use the multipole expansion to analyze the SCS of our nanostructure.
Figure~\ref{Fig2}a shows the multipole decomposition of the SCS based on Eq.\,\eqref{SCS}, where the full-wave SCS of Fig.\,\ref{Fig1}e, computed using Ansys Lumerical software package, is included for comparison.
One can see that only the ED contribution is sufficient to correctly describe the total SCS in the considered spectral range,  except for a narrow region around the minimum at $\lambda=735$ nm, where the MQ contribution is important, and the ED reaches its minimum corresponding to the anapole state. 
By applying Eq.~(\ref{LWA_p}), we can obtain the LWA decomposition of the exact ED  with inclusions of the quasi-static electric dipole ED$_0$ and the toroidal dipole TD. 
The result is  presented in Fig.\,\ref{Fig2}b.
One can see that the minimum of the ED at $\lambda=735$\,nm is mainly determined by the ED$_0$ contribution, whereas the TD provides only a weak ($\sim3$ nm) spectral shift to the minimum of ED$_0$. 
Moreover, away from the minimum, the contribution of the TD to the ED is smaller by several orders of magnitude compared to the ED$_0$ contribution. 
As a result, we could conclude that the realization of the dipole anapole state cannot be explained by simple destructive interference between the fields generated by the ED$_0$ and the TD. 
Indeed, this interference begins to play a role only when ED$_0$ is significantly suppressed. 
We see that the anapole state is already  formed due to this significant suppression of ED$_0$. 
Thus, in order to clarify the formation of the anapole state, we need to understand the reason for the suppression of ED$_0$. 
In the next section, we pursue a secondary multipole analysis to further investigate this issue.

\subsection{Secondary multipole analysis and anapole state}
To apply the secondary multipole analysis, we divide the total volume of the scatterer, $V$, into two parts, as shown in Fig.\,\ref{Fig2}e.
The first part, denoted by $V_1$, is the volume inside a cylinder positioned at the center of the nanostructure with a radius R$_1$ and an out-of-plane axis.
The second part, denoted by $V_2$, is the remaining volume outside the cylinder such that $V=V_1+V_2$. 
We then calculate the  total electric dipole moment, $\bf p$, of the nanostructure, and separately calculate the electric dipole moments ${\bf p}_1$ and ${\bf p}_2$ corresponding to $V_1$ and $V_2$, respectively.
The center of the nanostructure is considered the origin for all dipole moments. 
We use the vector equation ${\bf p}={\bf p}_1+{\bf p}_2$, where each dipole moment is calculated using Eq.\,\eqref{p0} and integrating over the corresponding volume.

The division of the nanostructure into two sub-volumes allows the representation of the total ED moment $\bf p$ in the form
\begin{equation}
    {\bf p}={\bf p}_1+{\bf p_2},
\end{equation}
where ${\bf p}_1$ (${\bf p}_2$) is the electric dipole moment corresponding to sub-volume $V_1$ ($V_2$).
By that, we can express the ED-SCS as
\begin{equation}\label{pp}
    \sigma_{\rm sca}^{p}=\frac{k_0^4}{6\pi\varepsilon_0^2|{\bf
E}|^2}\left\{|{\bf p}_1|^2+|{\bf p}_2|^2+2\Re({\bf p}_1\cdot{\bf p}_2^*)\right\}.
\end{equation}
By this representation, one can suggest another alternative to the realization of the anapole state without explicitly introducing the TD contribution. 
For example, given a wavelength $\lambda_{\rm anp}$ and sub-volumes $V_1$ and $V_2$, assume that the ED moments $|{\bf p}_1|= |{\bf p}_2|$ with both moments have a maximum value other than zero.
Then, we let $\sigma_{\rm sca}^p\to0$, which entails that the anapole state can be realized based on a destructive interference between the fields created by the ED moments ${\bf p}_1$ and ${\bf p}_2$.

Furthermore, the contribution of the TD to the formation of the anapole state can be estimated by using the expression
\begin{flalign}\label{pp0}
    \sigma_{\rm sca}^{p} &\approx \frac{k_0^4}{6\pi\varepsilon_0^2|{\bf
E}|^2}\left\{|{\bf p}_1^0|^2+|{\bf p}_2^0|^2+2\Re({\bf p}_1^0\cdot{\bf p}_2^{0*})\right.+\nonumber\\
&+\frac{k_d^4}{\omega^2}|{\bf T}|^2-2\frac{k_d^2}{\omega}\Im({\bf T}\cdot{\bf p}_1^{0*})-2\frac{k_d^2}{\omega}\Im({\bf T}\cdot{\bf p}_2^{0*})\}
\end{flalign}
which is the LWA of Eq.\,\eqref{pp}. 
In expression \eqref{pp0}, if the terms that include $\bf T$ are much smaller than $|{\bf p}^0_1|^2$ and $|{\bf p}^0_2|^2$, the TD would not play a principal role in the formation of the anapole state.
Instead, the anapole state could be formed by the interference of the quasi-static electric dipole moments ${\bf p}^0_1$ and ${\bf p}^0_2$.

Figure\,\ref{Fig2}c shows the contributions of the exact and the LWA dipole moments as functions of the radius R$_1$ at the wavelength $735$\,nm, where ${\bf p}$ is minimum.
Unlike ${\bf p}$, the magnitudes of ${\bf p}_1$ and ${\bf p}_2$ depend on the splitting radius R$_1$ and have essentially the same magnitude ($|{\bf p}_1|\approx|{\bf p}_2|$) with their maximum values occuring around R$_1=60$\,nm.
We note a similar behavior for ${\bf p}_1^0$ and ${\bf p}_2^0$ with a slightly noticeable difference between their magnitudes, which can be attributed to the weak contribution of the toroidal moment to the exact ED moment which also causes the red-shift of the dip in the scattering cross-section, see Fig.~\ref{Fig2}b.
It is worth mentioning that the computed phase difference $\angle{\bf p}_1- \angle{\bf p}_2\approx 180^\circ$, which explains their destructive interference shown in Fig.\,\ref{Fig2}c.

Since the ED moments are determined by the induced electric current, which is proportional to the total electric field in the structure, the maximum of ${\bf p}_1$ and ${\bf p}_2$, in Fig.\,\ref{Fig2}c, indicate that the electric field accumulates in $V_1$ and $V_2$ of the structure under the minimum of the total ED moment $\bf p$.
This analysis is consistent with our interpretation of the anapole state as a result of the destructive interference between the ED moments ${\bf p}_1$ and ${\bf p}_2$, associated with $V_1$ and $V_2$, respectively.
Figure~\ref{Fig2}d shows the amplitude of $|{\bf p}|$, $|{\bf p}_1|$, and $|{\bf p}_2|$ versus the wavelength when the nanostructure volume is divided using a radius R$_1$=60\,nm.
The minimum of $|{\bf p}|$ at $\lambda=735$~nm corresponds to the minimum of SCS (see Fig.\,\ref{Fig2}a) and does not depend on R$_1$, as demonstrated in Fig.\,\ref{Fig2}c.
Moreover, at the minimum, $|{\bf p}|<|{\bf p}_1|$ and $|{\bf p}|<|{\bf p}_2|$.
Such inequalities can only correspond to the destructive (out-of-phase) contributions  of ${\bf p}_1$ and ${\bf p}_2$ into the total $\bf p$.   
Moreover, Figs.\,\ref{Fig2}f-h show the field distribution inside the nanostructure for three wavelengths, at which the nanostructure exhibits the two peaks (at $580$\,nm and $985$\,nm) and the dip (at $735$\,nm) in the scattering cross-section.
In Fig.\,\ref{Fig2}f, the field lines are effectively pointing in the same downward direction in $V_1$ and $V_2$, which, given the subwavelength dimensions of the nanostructure, entails a constructive far-field interference, consistent with the results in Fig.\,\ref{Fig2}d around the wavelength $580$\,nm.
On the other hand, the field lines, in $V_1$ and $V_2$, have opposite orientations at $735$\,nm (Fig.\,\ref{Fig2}g) and at $985$\,nm (Fig.\,\ref{Fig2}h), which entails a destructive far-field interference which is consistent with the superposition of $\bf{p}_1$ and $\bf{p}_2$ in Fig.\,\ref{Fig2}d around the corresponding wavelengths.

Thus, the secondary multipole analysis and the field distribution confirm our conclusion about the formation of the anapole state by the destructive interference of two electric dipoles hosted by the nanostructure: one at the center and the other close to its external circumference. 
Note that the role of the TD moment for the realization of the anapole is not significant in this case, since the electric dipole moments ${\bf p}_1$ and ${\bf p}_2$ are mainly determined by the LWA ED$^0$ moments, ${\bf p}_1^0$ and ${\bf p}_2^0$, respectively, as follows from a comparison of the curves shown in Fig.\,\ref{Fig2}c.  
Indeed, the values of ${\bf p}_1$ and ${\bf p}_1^0$ are practically equal to each other, while the values of ${\bf p}_2$ and ${\bf p}_2^0$ are only slightly different
because of the weak contribution of the toroidal moment to the exact electric dipole moment ${\bf p}_2$.
Thus, the anapole state can be associated with the destructive (out-of-phase) interference of fields created by the LWA electric dipole moments ${\bf p}_1^0$ and ${\bf p}_2^0$ which have their maximum values around R$_1=60$\,nm.

\section{Metasurface building block}
\label{Sec:MetBuiBlo}

\begin{figure}
\centering
\includegraphics[trim = 0mm 6mm 0mm 2mm, clip,width=1.0\textwidth,draft=false,angle=0]{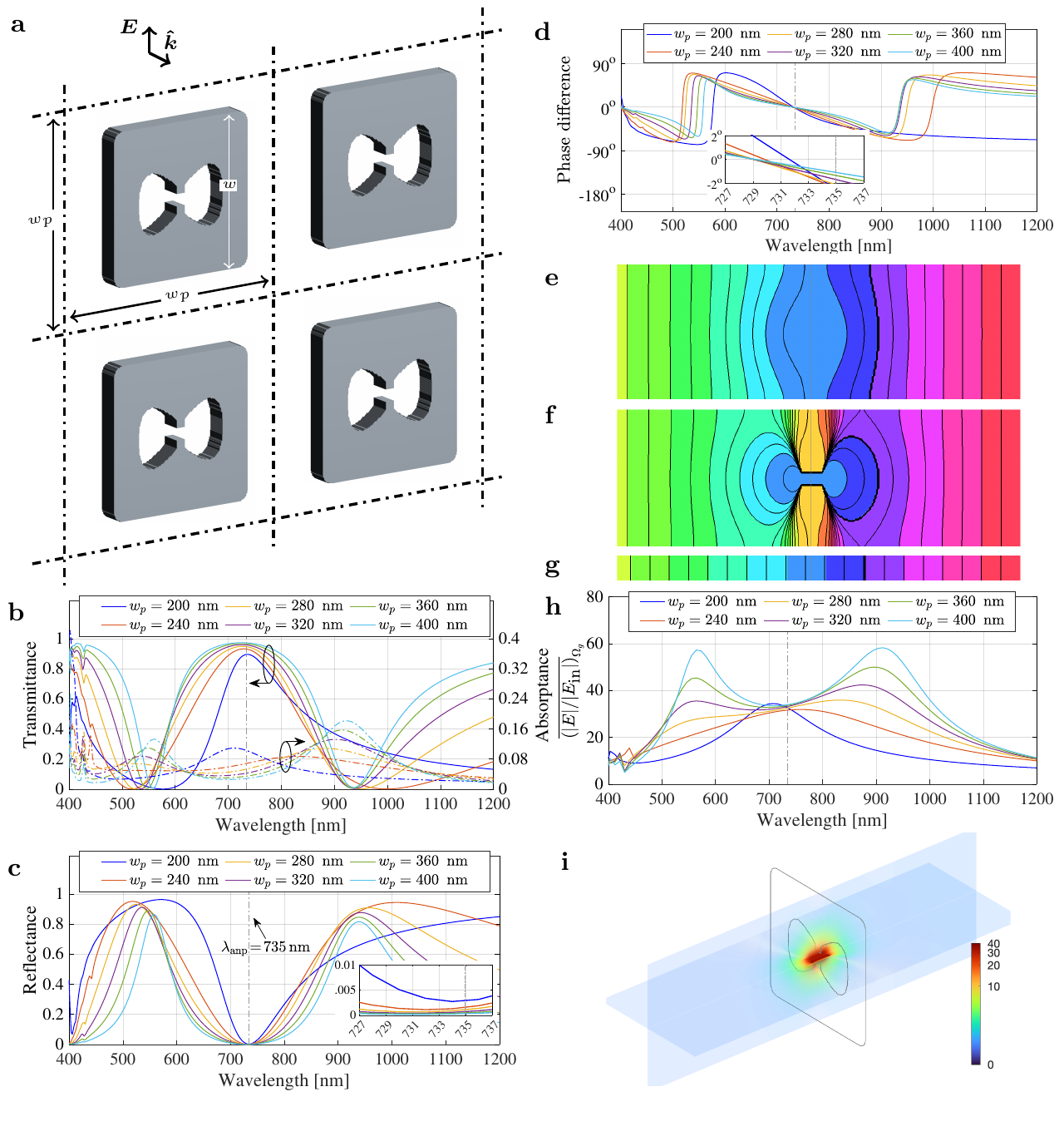}
\vspace{-0pt}
\caption{Anapole meta-atom as a building block for metasurfaces. (a) metasurface with a pitch $w_p$. (b) metasurfaces' transmittance (solid lines) and absorptance (dash-dotted lines).  (c) metasurfaces' reflectance. (d) difference in phase between the transmitted wave and an incident plane wave at a point sufficiently away from the nanostructure. The phase difference is zero  at\,$\lambda_{\rm anp}$. (e) and (f) are, respectively, the electric field phase profile at the vertical and the horizontal planes of symmetry for a unit cell with a pitch $w_p = 200$\,nm where $\lambda_{\rm anp} = 731.5$\,nm (see Visualizations 1). (g) the phase profile of the incident plane wave in free space, used as a reference.
(h) average field enhancement at the gap region. (i) electric field amplitude (slice view) for the unit cell with $w_p = 200$\,nm (see Visualizations 2).
}
  \label{Fig3}
   \vspace{-8pt}
\end{figure}

To prove that our plasmonic nanostructure acts as an anapole meta-atom, we need to test its robustness against electromagnetic coupling with adjacent nanostructures.
Thus, we use it as a building block for a metasurface, i.e., a 2D periodic arrangement of nanostructures, and investigate its performance versus the size of lattice periodicity, see Fig.\,\ref{Fig3}a.
The metasurface is simulated by using a square unit cell with a period size $w_p$ (henceforth, called the pitch), which we vary between $200$\,nm and $400$\,nm.
In simulations, a periodic boundary condition and a perfectly matched layer are enforced in the in-plane and the out-of-plane directions, respectively.
Note that $w_p= 200$\,nm corresponds to a zero separation between  the adjacent nanostructures of the metasurface (recall that the width of the nanostructure is $w=200$\,nm).
Fig.\,\ref{Fig3}b and Fig.\,\ref{Fig3}c show, respectively, the transmittance and the reflectance of the metasurface as a function of $w_p$.
Around the wavelength $735$\,nm and regardless of the pitch size, we note that all meatsurfaces exhibit essentially stable performances in terms of a vey low reflectance and a very high transmittance.
The metasurface exhbits a transmittance close to $97$\% for large $w_p$, and even for the extreme case, when the adjacent nanostructures have zero separation, the corresponding metasurface still maintains around $89$\% transmittance and a negligible reflectance around\,735\,nm.

Around the wavelengths $550$\,nm and $950$\,nm, all metasurfaces exhibit minima in transmittance which are correlated with maxima in the reflectance curves.
Since the resonant scattering of meta-atoms is determined only by their electric dipole moments, the metasurfaces exhibit the effect of total reflection at these resonances in accordance with the multipole transmission suppression theory \cite{babicheva2021multipole}.
As the pitch decreases, the spectral peaks/dips of the metasurface exhibit a blue-shift for wavelengths around $550$\,nm, a moderate red-shift for wavelengths around $950$\,nm, and essentially stable behavior around\,735\,nm.
The blue and red shifts of the peaks around $550$\,nm and $950$\,nm depend on the lattice pitch $w_p$ and can be explained as a result of near-field interaction of adjacent resonant nanostructures~\cite{rechberger03optical}.
When the adjacent nanostructures touch each other (i.e. when $w_p=200$), we observe noticeable abrupt changes in the spectral peaks/dips at the wavelengths around $550$\,nm and $950$\,nm, but not for the peaks/dips around $735$\,nm.
Fig.\,\ref{Fig3}b shows that the absorptance of the unit cell increases as $w_p$ decreases, with a maximum value of $\sim\!11$\% for the extreme case $w_p=200$\,nm and $\sim\!3$\% when $w_p=400$\,nm.
On the contrary, around the $550$\,nm and $950$\,nm, the unit cell absorptance increases as $w_p$ increases, which is correlated with the increase in the field enhancement, given in Fig.\,\ref{Fig3}h.
That is the absorptances of the resonant electric dipoles and the anapole state exhibit opposite behavior regarding the dependency on the pitch parameter $w_p$.

Figure\,\ref{Fig3}d shows the phase difference between the transmitted wave and the incident plane wave at a distance sufficiently away from the nanostructure.
The figure shows that around the wavelength $735$\,nm and regardless of the pitch size, all metasurfaces contribute essentially a zero phase delay.
The inset in the same figure shows that the wavelength at zero-crossing phase (henceforth, referenced as $\lambda_{\rm anp}$) depends slightly on $w_p$.
For large values of $w_p$,  $\lambda_{\rm anp}$ converges to $ \simeq 729$\,nm. 
That is, a $6$\,nm blue-shift compared to the anapole wavelength $735$\,nm of the single nanostrucutre in free space. 
As $w_p$ decreases, the value of $\lambda_{\rm anp}$ exhibits a red-shift with a maximum value of  $\lambda_{\rm anp} \simeq 731.5$\,nm at $w_p= 200$\,nm, that is when adjacent nanostructures touch each other. 
Also, we note that the curves become steeper as $w_p$ decreases with a maximum phase change of $2^\circ$ between $729$\,nm and $731.5$\,nm, indicating a very small dependency of $\lambda_{\rm anp}$ on the pitch parameter. 
Figure\,\ref{Fig3}e and Fig.\,\ref{Fig3}f show the phase distribution of the electric field across the vertical and horizontal planes of symmetry of a unit cell with $w_p= 200$\,nm at $\lambda_{\rm anp}=731.5$\,nm.
To the right of the metasurface where the transmitted wave is monitored, we see a flat wavefront, represented by the vertical lines in Fig.\,\ref{Fig3}e and Fig.\,\ref{Fig3}f.
The wavefront is aligned with that of a plane wave propagating in free space (see Fig.\,\ref{Fig3}g).
To the left of the metasurface, we observe a tiny misalignment  which we attribute to the finite reflectance (see the inset in Fig.\,\ref{Fig3}c) that causes interference with the incident wave.
We notice that the phase at the gap location matches the phase of the unperturbed plane wave, thus creating an opportunity for phase control at the nanoscale.
These results demonstrate the transparent character of metasurfaces built by using our nanostructure, around the anapole wavelenght\,$\lambda_{\rm anp}$.

Figure\,\ref{Fig3}h shows the average field enhancement at the gap region of the unit cell for different values of $w_p$. 
Around the wavelengths $550$\,nm and $950$\,nm, the field enhancement is highly dependent on $w_p$, which could be attributed to the near-field interaction between the adjacent meta-atoms.
As $w_p$ decreases, the increase of the mutual interaction deteriorates the field enhancement of the individual nanostructures. The field enhancement in the gap depends on $w_p$ except around\,$\lambda_{\rm anp}$, where the metasurface unit-cell exhibits essentially the same enhancement capability as the single element in free space (see Fig.\,\ref{Fig1}d).
Fig.\,\ref{Fig3}i shows a slice view demonstrating the electric field enhancement in the gap at the anapole wavelength $\lambda_{\rm anp}=731.5$\,nm, for the unit cell with a pitch $w_p=200$\,nm.

In summary, the anapole state possessed by the single nanostructure is maintained when it is used to build metasurfaces. 
We notice only tiny shifts of\,$\lambda_{\rm anp}$ that, for the extreme case, doesn't exceed 6\,nm.
At the anapole state, all metasurfaces exhibit high transmittance, negligible reflectance, and high field enhancement.
At the anapole state, the phase of the transmitted wave is identical to the incident wave, so that for configurations with very weak absorption, the effect of the metasurface transparency is realized.
To emphasize the behaviour as a meta-atom, we also provide further evidence of robustness against different surrounding conditions.
In Appendix\,\ref{subsec:ImpSur}, we demonstrate that the anapole state is preserved also in presence of a substrate or a homogeneous background different than air with only a spectral shift of the anapole wavelength.
In addition, we demonstrate that the anapole state is polarization dependent (see Appendix\,\ref{subsec:ImpPol}), and an optical switching mechanism is observed between high and low transmittance for vertical ($z$-axis) and horizontal ($y$-axis) polarization, respectively.

 \section{Application to metamaterials}
 \label{Sec:AppMet}
 
 \begin{figure}
\centering
\includegraphics[trim = 0mm 5mm 0mm 0mm, clip,width=0.96\textwidth,draft=false,angle=0]{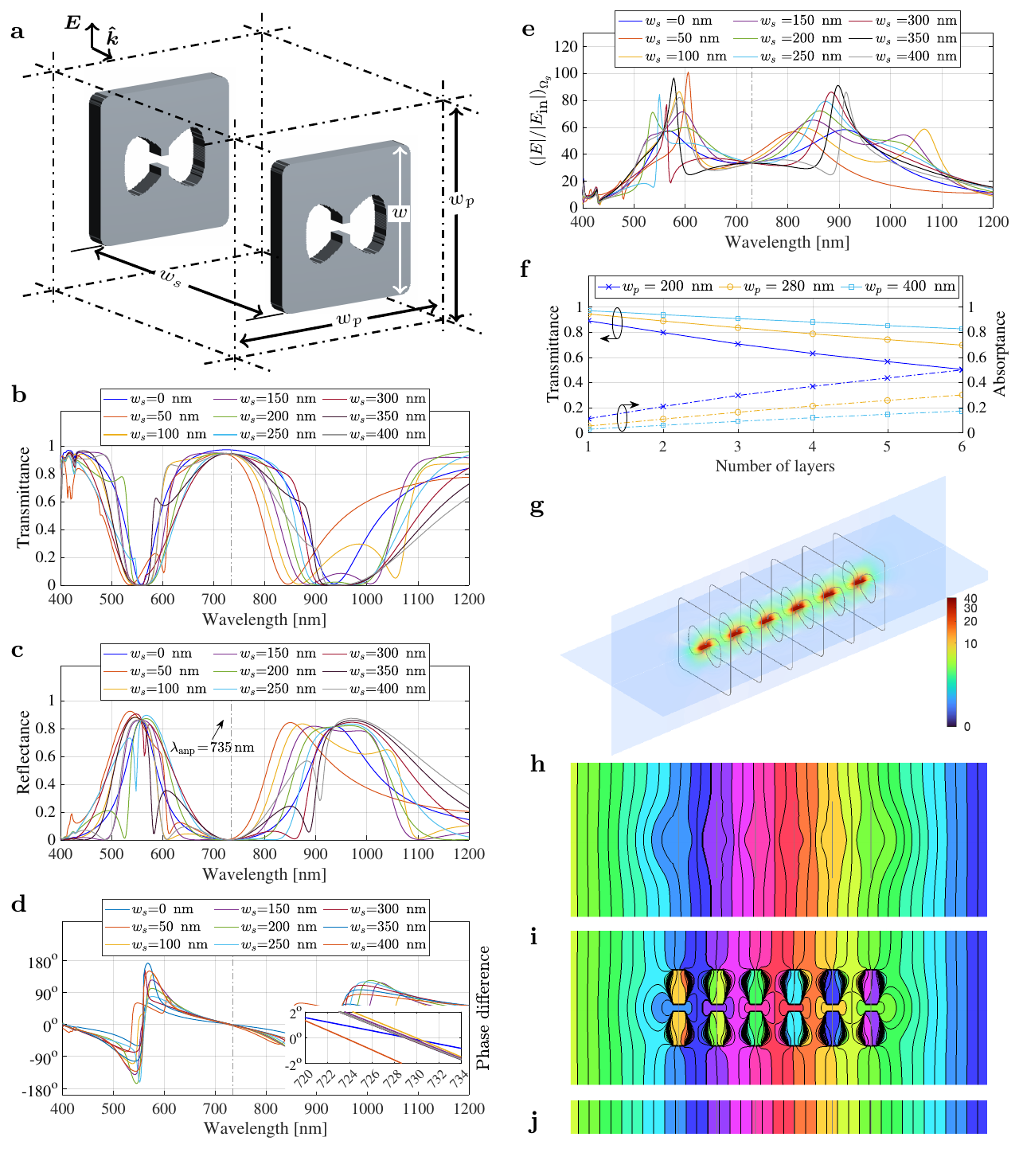}
\vspace{-0pt}
\caption{Anapole-based metamaterials. (a) geometrical parameters of a two-layer unit cell of the metamaterial. (b) the unit cell transmittance versus $w_s$ with $w_p=400$.  (c) the reflectance. (d) the difference in phase between the transmitted wave and an incident plane wave at a point sufficiently away from the nanostructure. The phase difference is zero at the anapole wavelength. (e) Average field enhancement at the gap region of the first layer. (f) transmittance and absorptance versus the number of layers for three pitch values $w_p$ and a fixed $w_s=100$\,nm. (g) A slice view showing the electric field amplitude over the unit cell of a 6-layer metamaterial with $w_p=400$\,nm and $w_s=100$\,nm at $\lambda_{\rm anp}=730$\,nm (see Visualizations 3).  (h) and (i) are, respectively, the electric field phase distribution in the vertical and horizontal planes of symmetry of the 6-layer unit cell in (g) (see Visualizations 4). (j) the phase distribution of a $730$\,nm incident plane wave, used as a reference.}
  \label{Fig4}
   \vspace{-8pt}
\end{figure}

The transparency enabled by our optimized anapole meta-atom in metasurfaces raises the question whether, and to what extent, transparency would be preserved in a metamaterial configuration.
In this section, we study the characteristics of metamaterials constructed using our anapole meta-atom.
First, we investigate thin metamaterials comprising two-layer metasurfaces, see Fig.\,\ref{Fig4}a.
Second, we extend our study to thick metamaterials that involve multilayer metasurfaces.
Initially, we fix the pitch to $w_p = 400$\,nm, then we provide results for other values of $w_p$.

Figures\,\ref{Fig4}b-e show the transmittance, reflectance, phase delay, and the field enhancement of a two-layer metamaterial as a function of the separation distance, $w_s$, between the layers.
The case $w_s=0$\,nm marks the performance of the single layer metasurface with $w_p = 400$\,nm and is used as a reference.
Compared to the single metasurface, we note that the performance of the metamaterial around the anapole state is maintained for separation distances larger than $150$\,nm, see the inset in\,Fig.\,\ref{Fig4}d, where $\lambda_{\rm anp}$ converges to the $729$\,nm of the single-layer metasurface (see the inset in Fig.\,\ref{Fig3}d).
Also, at the anapole wavelength, the transmittance amplitude decreased only by $3$\% compared to the single-layer metasurface as shown in Fig.\,\ref{Fig4}b.
Figure\,\ref{Fig4}c shows, similar to the single-layer metasurface, that the reflectance amplitude is negligible around the anapole wavelength.
Figure\,\ref{Fig4}d shows the phase delay of the two-layer metamaterial for different values of $w_s$. 
The two-layer metamaterial possesses essentially the same value of $\lambda_{\rm anp}=729$\,nm of the single layer metasurface (i.e. $w_s=0$\,nm) when the separation distance $w_s$ is greater than $150$\,nm.
Figure\,\ref{Fig4}e shows the field enhancement (at the gap of the first layer) of the two-layer metamaterial, which emphasizes the unperturbed performance around the anapole wavelength. 

We extend our investigations to thick metamaterials.
Here, we choose $w_s=100$\,nm and append metasurfaces in the direction of propagation to form the metamaterials.
Figure\,\ref{Fig4}f summarizes the characteristics of metamaterials with different thicknesses for three values of the pitch parameter $w_p$.
For comparison, we include the single metasurface for each case (i.e., when the number of layers is one).
The average attenuation per layer is 7.7\%, 5.0\%, and 2.9\% for $w_p$ equals $200$\,nm, $280$\,nm, and $400$\,nm, respectively.
The highest attenuation rate in transmittance corresponds to the case when $w_p=200$\,nm, that is, when the anapole state of the nanostructure experiences the maximum disturbance from adjacent nanostructure in the lattice.
Also included in Fig.\,\ref{Fig4}f are the absorptance curves for the three values of $w_p$.
The maximum reflectance (not presented here) is below  0.6\% for all cases, which entails that the attenuation in transmittance is mainly caused by the absorption in the nanostructures.
We note that as the pitch size increases, the absorptance rate decreases.

Figure\,\ref{Fig4}g shows a slice view for the electric field distribution of the unit cell of a 6-layer metamaterial with a pitch $w_p=400$\,nm and interlayer separation $w_s=100$\,nm.
Figure\,\ref{Fig4}h and Fig.\,\ref{Fig4}i show the phase of the electric field of the 6-layer unit cell at $\lambda_{\rm anp}=730$\,nm, and Fig.\,\ref{Fig4}j depicts the corresponding phase profile of a plane wave propagating in free space.
The results demonstrate the transparency of the nanostructure when used to build thick metamaterials with as little as 3\% per layer attenuation possible to achieve by choosing proper values of the pitch $w_p$ and the interlayer separation $w_s$ parameters.
We believe that a plethora of applications could benefit from the hot spots developed in the gap of each nanostructure along the propagation direction. Furthermore, the phase at the gap locations matches the phase of the reference plane wave, thus creating an opportunity for coherent emission if molecules or nonlinear materials are placed in the gaps.

\begin{figure}
\centering
\includegraphics[trim = 0mm 2mm 0mm 0mm, clip,width=1.0\textwidth,draft=false,angle=0]{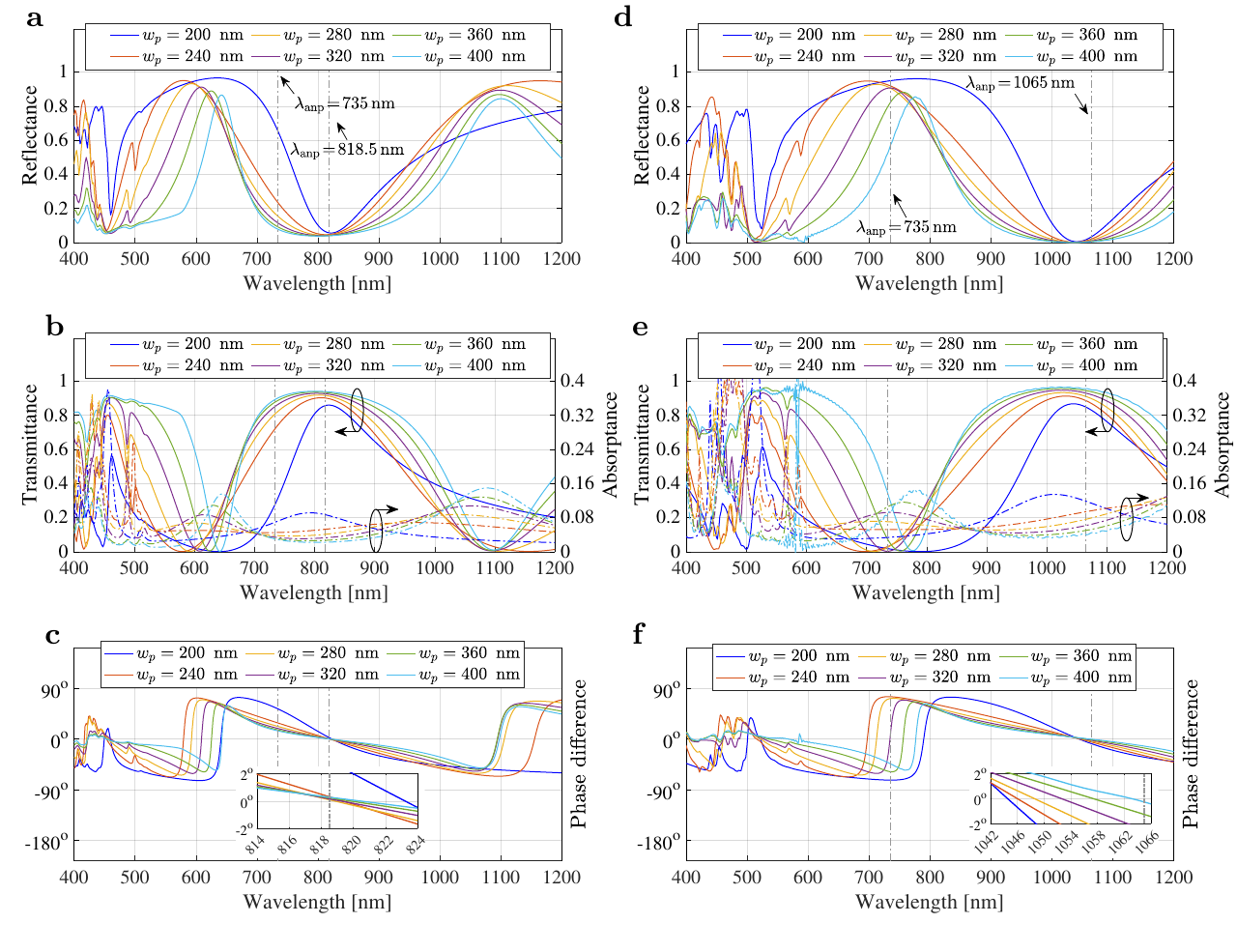}
\vspace{-8pt}
\caption{Performance of the single-layer metasurface when: 1) placed on top of a glass substrate with refractive index $1.45$. (a) reflectance; (b) transmittance; and (c) phase difference between transmitted wave and a plane wave in free space. The anapole wavelength redshifts to $\sim 818.5$\,nm. 2) (d)-(f) are the corresponding performance when the metasurface is immersed in a background glass medium. The anapole wavelength redshifts to $\sim 1065$\,nm. The anapole state is maintained, however, the wavelength is tuned by the refractive index of the surrounding medium.}
  \label{MetasurfaceSub}
   \vspace{-8pt}
\end{figure}

\section{Conclusion}
\label{Sec:Con}
We theoretically and numerically demonstrated the achievement of a near-ideal anapole state in the visible regime by means of a thin planar plasmonic nanostructure. The nanostructure exhibits a very low extinction cross-section, which enables transparency, and a field enhancement in the gap volume of the structure while maintaining low losses. Such design was obtained via topology optimization while optimizing for broadband field enhancement. 
Applying multipole analysis, we got new insight into the formation of the anapole state based mainly on the destructive interference of two quasi-static electric dipole moments associated with different parts of the structure. 
The anapole state is maintained even when the nanostructure is used in highly packed metasurfaces or metamaterials, including the case when placed on a substrate. These combined features make our nanostructure act as an independent meta-atom.
The field confinement outside the nanostructure enables applications in sensing and laser cavities, while the high density of plasmonic hot-spots with the phase in the gap matching the phase of the reference plane wave can be beneficial for nonlinear light enhancement and control. 
Furthermore, the planarity of the design
makes the manufacturing simpler than previously reported plasmonic anapoles.
Our study, also based on the predicted low losses at the anapole state, could open new opportunities for plasmonics, as most of the features of our design are not achievable by dielectric structures with similar dimensions. 
Finally, we remark the huge potential of inverse design techniques to unlock new physics and functionalities for light-matter interaction at the nanoscale.

\section{Appendix}
\label{sec:App}

\subsection{Impact of surrounding medium}
\label{subsec:ImpSur}

\begin{figure}
\centering
\includegraphics[trim = 0mm 1mm 0mm 0mm, clip,width=1.0\textwidth,draft=false,angle=0]{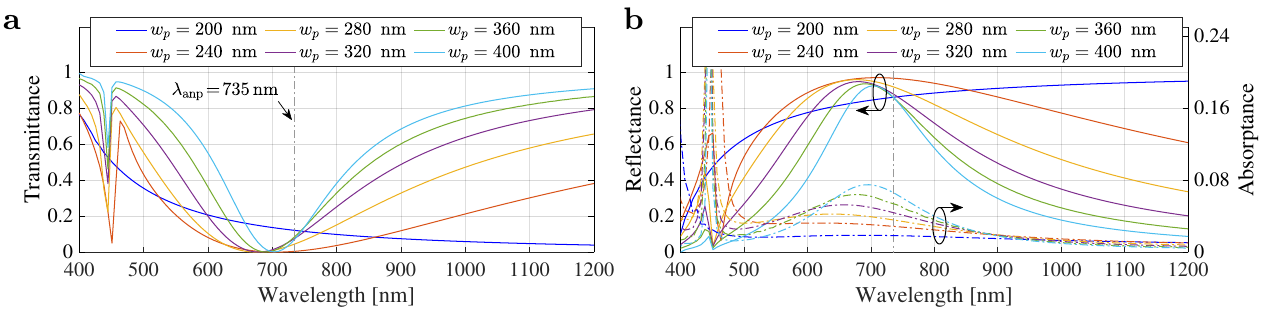}
\vspace{-8pt}
\caption{Performance of single-layer metasurfaces when illuminated with an incident plane wave polarized in the orthogonal direction ($y$-direction in Fig.\,\ref{Fig1}), for different values of the pitch parameter $w_p$. (a) transmittance.  (b) reflectance (solid lines) and absorptance (dash-dotted lines). The case $wp=240$\,nm exhibits more than $96$\% reflectance and less than $0.5$\% transmittance, demonstrating that the metasurfaces can be utilized as high-performance polarizers.}
  \label{MetasurfaceXPolPeriodicity}
   \vspace{-8pt}
\end{figure}
In this section, we study the influence of the surrounding medium on the anapole state.
Figure\,\ref{MetasurfaceSub}a-c show the performance of single-layer metasurfaces when placed on top of a glass substrate with a 1.45 refractive index.
Compared to the free space, we observe two changes. 
Firstly, the anapole wavelength redshifts to the wavelength around $818$\,nm for large values of $w_p$.
Secondly, the amplitude of the reflectance/transmittance curves is slightly increased/decreased by $3.4$\%, which is almost the same amount that a plane wave experiences at normal incidence on an air-glass interface.
That is, the anapole state is maintained when the metasurface is placed on a glass substrate with only a red-shift of the anapole wavelength.
Figure\,\ref{MetasurfaceSub}d-f shows the metasurfaces' performance when surrounded by a  background glass medium.
The anapole wavelength redshifts from $\sim735$\,nm to around $735\times 1.45 = 1065$\, nm.
That is, the wavelength of the anapole is proportional to the refractive index of the surrounding medium.
In this case, we observe a noticeable sensitivity of the zero-crossing wavelength $\lambda_{\rm anp}$ to the metasurface pitch parameter $w_p$, see the inset in Fig.\,\ref{MetasurfaceSub}f.

\subsection{Impact of polarization}
\label{subsec:ImpPol}
Here, we investigate the performance of the metasurfaces when excited with orthogonal polarization (i.e. when the  polarization of the incident wave is rotated by $90^\circ$ relative to the sketch in Fig.\,\ref{Fig3}a).
Figure\,\ref{MetasurfaceXPolPeriodicity}a shows the transmittance and the reflectance when the metasurface is excited with the orthogonal polarization.
The metasurfaces exhibit low transmittance and high reflectance at the wavelength $735$\,nm.
The dip in transmittance exhibits a small blueshift as $w_p$ decrease with abrupt change occurring when adjacent nanostructures touch each other (i.e. when $w_p=200$\,nm).
At the wavelength $735$\,nm and for the selected range of $w_p$, the transmittance is below $8.5$\% and the reflectance is above $86$\% as shown in Fig.\,\ref{MetasurfaceXPolPeriodicity}a and Fig.\,\ref{MetasurfaceXPolPeriodicity}b.
When $w_p = 240$\,nm, the metasurface exhibits a broadband performance, where the transmittance is below  $0.5$\% and the reflectance is above $96.5$\%.
The difference between the transmittance and the reflectance is the losses absorbed by the metasurfaces, which is shown in Fig.\,\ref{MetasurfaceXPolPeriodicity}b.
Under orthogonal irradiation, the metasurfaces exhibit a high optical switching mechanism that can be utilized, for instance, to build optical polarizers.
 
\section*{Acknowledgement}
The computations were performed on resources provided by the Swedish National Infrastructure for Computing (SNIC) at HPC2N center; the central computing cluster operated by Leibniz University IT Services (LUIS); and the North German Supercomputing Alliance (HLRN) as part of the NHR infrastructure. A.C.L. acknowledges the German Federal Ministry of Education and Research (BMBF) under the Tenure-Track Program.  A.B.E. and A.C.L. acknowledges the support of the Deutsche Forschungsgemeinschaft (DFG, German Research Foundation) under Germany’s Excellence Strategy within the Cluster of Excellence PhoenixD (EXC 2122, Project ID 390833453). 
The publication of this article was funded by the Open Access Fund of the Leibniz Universität Hannover.


\section*{Funding}
The Deutsche Forschungsgemeinschaft (DFG, German Research Foundation) under Germany’s Excellence Strategy within the Cluster of Excellence PhoenixD (EXC 2122, Project ID 390833453).

\providecommand{\latin}[1]{#1}
\makeatletter
\providecommand{\doi}
  {\begingroup\let\do\@makeother\dospecials
  \catcode`\{=1 \catcode`\}=2 \doi@aux}
\providecommand{\doi@aux}[1]{\endgroup\texttt{#1}}
\makeatother
\providecommand*\mcitethebibliography{\thebibliography}
\csname @ifundefined\endcsname{endmcitethebibliography}
  {\let\endmcitethebibliography\endthebibliography}{}

\end{document}